\documentclass[journal]{IEEEtran}
\usepackage{comment}
\usepackage{amsmath,amssymb,bm}
\usepackage{tabularx}
\usepackage[english]{babel}

\usepackage{pgfplots}
\usepackage{graphicx}
\usepackage{booktabs} 
\usepackage{caption}
\usepackage{lscape} 
\usepackage{subcaption}
\usepackage{xcolor, soul, color}
\usepackage{cite}
\usepackage{bbm}
\usepackage{colortbl} 
\usepackage{algorithmic}
\usepackage[linesnumbered,ruled,vlined]{algorithm2e}
\SetKwComment{Comment}{\tcp*{}}{}
\SetKwInOut{KwIn}{Input}
\SetKwInOut{KwOut}{Output}
\SetKwInOut{KwData}{Initialize}
\usepackage[colorlinks=true,  linkcolor=black, citecolor=blue, breaklinks=true, urlcolor=blue]{hyperref}

\definecolor{customgray}{HTML}{F5F5F5}

\usepackage{tabularx}
\usepackage{pgfplots}
\usepackage{tikz}
\usepackage[font=small,skip=5pt]{caption}
\usepgfplotslibrary{groupplots}
\pgfplotsset{compat=1.13}
\newlength\fwidth

\newcommand{\nonl}{\renewcommand{\nl}{\let\nl\oldnl}}

\ifCLASSINFOpdf
\else
\fi

\begin{document}
\bstctlcite{IEEEexample:BSTcontrol} 
\title{Meta-Learning Based Radio Frequency Fingerprinting for GNSS Spoofing Detection}
\author{\IEEEauthorblockN{Leatile Marata, \textit{Member, IEEE} Juhani Sankari, \textit{Member, IEEE},  Eslam Eldeeb, \textit{Member, IEEE},  Mikko Valkama, \textit{Fellow, IEEE}, Elena Simona Lohan, \textit{Senior Member, IEEE}}
\thanks{Leatile Marata, Juhani Sankari, Elena Simona Lohan and Mikko Valkama are with Tampere Wireless Communications Centre, Tampere University, Finland. e-mail: \{leatile.marata,Juhani.Sankari,simona.lohan, mikko.valkama\}@tuni.fi. Eslam Eldeeb is with the Centre for Wireless Communications Centre, University of Oulu, Finland. e-mail: Eslam.Eldeeb@oulu.fi}
\thanks{This work was supported by the Research Council of Finland under RESILIENT project (number 359846).}} 
\maketitle
\begin{abstract}
 The rapid development of technology has led to an increase in the number of devices that rely on position, velocity, and time (PVT) information  to perform their functions. As such, the Global Navigation Satellite Systems (GNSS) have been adopted as one of the most promising solutions to provide PVT. Consequently, there are renewed efforts aimed at enhancing GNSS capabilities to meet emerging use cases and their requirements. For example, GNSS is evolving to rely on low-earth-orbit satellites, shifting the focus from traditional medium-earth-orbit satellites. Unfortunately, these developments also bring forth higher risks of interference signals such as spoofers, which pose serious security threats. To address this challenge, artificial intelligence (AI)-inspired solutions are being developed to overcome the limitations of conventional mathematics-based approaches, which have proven inflexible when dealing with diverse forms of interference. In this paper, we advance this direction by proposing a meta-learning framework that enables GNSS receivers to detect various types of spoofers. Specifically, our approach exploits the radio frequency fingerprints present in the signal at both the pre-correlation and post-correlation stages of the receiver. The proposed solution has superior generalization properties compared to the state-of-the-art solutions. Numerical results demonstrate that our proposed solution significantly detects spoofers of different forms, with spoofing detection accuracies of more than $95\%$ on multiple datasets from the Texas Spoofing Test Battery (TEXBAT) and the Oak Ridge Spoofing and Interference Test Battery (OAKBAT)  repositories.
\end{abstract}

\begin{IEEEkeywords} Global Navigation Satellite Systems (GNSS), Model-Agnostic Meta-Learning (MAML), OAKBAT, Radio Frequency Fingerprinting (RFF), TEXBAT, Spoofing
\end{IEEEkeywords}

\section{Introduction}
\label{Introduction}
The evergrowing demand for a fully connected society and industrial automation is primarily supported by interconnected devices, such as sensors and actuators, that operate with minimal human intervention. To fully benefit from these systems, it is sometimes necessary for both the devices and their users to be aware of their environment. This awareness can be enabled through position, velocity, and time (PVT) parameters. For instance, the envisioned self-driving vehicles require PVT solutions for navigational purposes as well as to avoid collisions \cite{8998218,9751089,MARATA2019389}. It is worth noting that the Global Navigation Satellite Systems (GNSS) is still among the most widely used system for providing such PVT solutions \cite{ 9937069}. As a result, GNSS is currently developing at a rapid pace to continue providing solutions for existing services, as well as to adapt to new service requirements. Some of these developments include the modernization of existing Medium Earth Orbit (MEO) satellites used for GNSS and the development of GNSS supported by Low Earth Orbit (LEO) satellites. Unfortunately, there has also been a rapid development of Software-Defined Radios (SDRs), which facilitate the inexpensive creation of interference signals that can pose both security and availability threats to GNSS. These interferers can take the form of jammers (which block genuine signals) and spoofers (which mimic real signals). While both are detrimental to GNSS, spoofers generally pose a greater security threat, such as in air navigation, because they can alter the PVT fix \cite{elghamrawy2021high}. Owing to this, spoofer signal detection solutions have been a subject of study over the years as we discuss in the next paragraph. 

\par A lot of work on spoofing detection approaches has been ongoing. Some of these solutions include Signal Quality Monitoring (SQM) at the precorrelation phase, the tracking phase, and during navigation data processing \cite{zidan2020gnss,Sun2025}. These methods use signal-processing techniques to analyze parameters such as power levels, signal time of arrival, and correlator output in the receiver. It is important to note that such solutions rely on modeling the signal based on certain assumptions about the true/genuine signal and about the behaviour of the spoofer, and are therefore only as accurate as the underlying model. However, a well-known drawback of such mathematical models is their limited flexibility to adapt. For instance, in the case of complex spoofing attacks, models developed under the assumption of simple spoofers often fail \cite{10495074}.

\par To overcome the limitation of mathematics-inspired spoofing-detection solutions, a lot of work has been put towards the developement of artificial intelligence (AI)-inspired solutions for GNSS receivers.
Some of these AI-inspired solutions heavily rely on exploiting certain features that are highly vulnerable to their preprocessing stage and have limited performance under limited data. A newer approach, which is of particular interest in this paper involves the use of Radio Frequency Fingerprinting (RFF) to detect the presence or absence of a spoofer \cite{s24237698}. It is worth noting that the Radio Frequency (RF) fingerprints are unique hardware impairments inherent to transmitter devices, even those manufactured according to the same specifications \cite{10938874}. For this reason, if RF fingerprints can stored beforehand, they could offer a promising paradigm for detecting different signal sources and making distinction between these sources, such as those from SDR and those from genuine GNSS satellites. 

\par Motivated by this, our work develops a novel meta-learning based receiver algorithm that is aimed at detecting spoofers in GNSS, by combining different data from different stages of the receiver, which has received little attention so far. Our work also differs from other studies, such as \cite{s24237698,9847540}, as we extract features from the precorrelation stage using spectrograms and rely on a meta-learning framework, which enhances the transferability of the learned features. Next, we first present the state of the art that has influenced the development of our work, then we emphasize this paper's contributions. 
\subsection{State of the Art}
\label{STATEOFART}
 In general, the learning-based approaches are designed to mimic the thinking capabilities of a human being, which ultimately makes them adaptable and scalable. Consequently, applying such learning tools in areas such as spoofer detection, which is the main focus of our work, has also received considerable attention. These solutions can be classified into supervised learning and unsupervised learning. For instance, Bose \textit{et al.} in \cite{9656691} proposed a Neural Network (NN) Machine Learning (ML) solution for detecting spoofers. In \cite{9656691}, the authors mainly used the carrier to noise ratio (CNR), the Doppler shift and carrier phase data. The results showcased a detection accuracy of around $98\%$ when using three different features for classification. As such, the proposed solution does not consider the precorelation data. Moreover, the data sets used in \cite{9656691} is not publicly available, hindering the reproducibility of the results. Also, no cross-testing aspects was considered. By cross-training we refer to the generalizability property of a ML algorithm to achieve good results on testing datasets completely different from the training datasets (e.g., collected under different conditions and spoofing parameters). 
 
 \par  Another relevant work is \cite{9844986} by Chen \textit{et al.}, where the authors proposed a multi-parameter-based spoofing detection solution using a Support Vector Machine (SVM) and multiple post-correlation parameters from the GNSS receiver. Their results, studied on several datasets from Texas Spoofing Test Battery (TEXBAT) and the Oak Ridge Spoofing and Interference Test Battery (OAKBAT)  repositories, showed that the spoofing detection capabilities improve with the number of features, as demonstrated by an increase in accuracy from $76\%$ when using a single parameter to $96\%$ when using seven parameters. The proposed solution in \cite{9844986} fully relied on post-correlation data and no crosss-testing/generalizability aspects are considered.
  Dang et al. in \cite{9845684} proposed a learning-based GPS anti-spoofing solution for unmanned aerial vehicles (UAVs), using UAV trajectory deviation as a feature, thus relying on navigation data. Their work demonstrates a spoofer detection accuracy of $97\%$. However, this performance is highly reliant on the number of base stations (BS). In \cite{9456965} Shafique \textit{et al.} proposed a spoofing detection mechanism in GPS for UAV by using a mechanism of selecting different ML models based on performance. Some of the features used in this work include jitter, shimmer, and frequency modulation parameters. The proposal achieves an accuracy of approximately $99\%$ as well as an F1-score of $0.98$. Similar to \cite{9844986,9845684}, this work focuses on postcorrelation data and not precorelation. 
 
\par Pardhasaradhi \textit{et al.} in \cite{9925146} also proposed an ML algorithm that uses the received power and the correlation-function distortion features to screen the presence and absence of the spoofing signal in GNSS receiver. The results tested accross various ML algorithms such as SVM and neural networks  showcase test accuracies of around $98\%$. Unfortunately, this work was also not tested on publicly available datasets, thus it has limited reproducibility value. In a same vein, Yang \textit{et al.} \cite{10109166} proposed a reinforcement-learning solution for detecting spoofers by learning trajectory patterns to identify spoofing activities in autonomous vehicles. Reinforcement learning is an ML approach that allows the learning agent to interact with the environment, which in this case is the trajectory patterns \cite{nguyen2020deep}. The results from their work showed an improved detection capabilities on real data sets, but again,  the datasets used in \cite{nguyen2020deep} are not publicly available. 

One of the outstanding limitations of the above-mentioned works, is that they are not studied in terms of their cross-testing or generalizability capability for many cases of spoofing scenarios. In a quest for a more general solution, Iqbal \textit{et al.} proposed a deep-learning spoofing detection algorithm in \cite{10495074}, specifically using variational autoencoder to achieve generalization. Note that variational autoencoders typically learn the distribution of the training data so as to regenerate similar data. As such, an algorithm of this nature has the ability to generalize to detect different forms of spoofers \cite{wei2020recent}. In the work of \cite{10495074}, the results demostrated a high classification accuracy of more than $99\%$ for Texas Spoofing Test Battery (TEXBAT) repository, which is publicly available, but only with similar datasets for testing and training. In terms of generalizing performance to other data sets, much lower accuracies were achieved, for example a spoofing detection accuracy of only $44\%$ for TEXBAT DS7 dataset. Moreover, general algorithms of this nature require large amounts of training data, which is sometimes impractical to have. 

\par  All in all, there have been several  AI-inspired works that attempted to address the spoofing detection in GNSS receiver.  Table \ref{TableComparison} provides an overview at a glance of the related works based on ML for spoofing detection in GNSS.
However, as noted from the discussion in this subsection as well as from Table \ref{TableComparison}, most of these  solutions are not generalized to function on GNSS receivers in diverse spoofing conditions.  For instance, models that are trained on simple spoofing scenaorios cannot be generalized to complex spoofing scenarios. This has been the case for several works \cite{10495074,Marata2025} studying cross-testing or generalizability aspects with the TEXBAT data, including eight public datasets, named as  $\text{ds}_{i}$, $i=0,\cdots, 8$. As such, the problem of general spoofing detection requires novel solutions. Furtunately, in other areas of communications such as wireless there are some ongoing works, such as those that use Meta-Learning algorithms \cite{Li2021,Yu2023} to achieve generalization in classification tasks. This approach has been proven to provide better transferability of features accross different data sets \cite{10948463}. For example, the  works in\cite{9428530,vilalta2002perspective,10413635} are valuable resources for providing background on Meta Learning. Motivated by this, our current work focuses on developing a spoofing-detection framework that can work for different spoofing types by leveraging the Meta Learning framework. 

To the best of our knowledge of the literature, there has not been any works that use meta learning for spoofing detection, thus opening new research avenues in GNSS.  The contributions of our work are summarized in the next subsection. 
\begin{table*}[t!]
 \caption{A comparative study of different ML approaches that use the RF fingerprinting technique}
    \centering
    \footnotesize{
    \renewcommand{\arraystretch}{1.3} 
    \setlength{\arrayrulewidth}{1pt} 
    \begin{tabular}{|>{\columncolor{customgray}}p{2.0cm}|>{\columncolor{customgray}}p{4.6cm}|>{\columncolor{customgray}}p{3.4cm}|>{\columncolor{customgray}}p{5.8cm}|}
        \hline
        \textbf{Reference and year} & \textbf{Approach} & \textbf{Datasets} &  \textbf{Performance Analysis} \\ 
        \hline
        Wang \textit{et al.}\cite{9847540}, 2023&  Classification based on support vector machine (SVM) & Data generated from Tampere, Nottingham and Nuremberg (UK)& The proposed solutions achieves $99.99\%$ for spectrogram classification and $87,72\%$ for precorelation classification. The work does 
 not exploit corelation between precorelation and postcorelation phases. No cross-testing. \\ 
        \hline
        Zhang \textit{et al.}\cite{10679860}, 2024 & Classification based on various RF feautures using autoencoder & Dataset generated inhouses  & Performance of around $99\%$ for specific data set and not tested on publicly available data sets.  No cross-testing.  \\  
        \hline
        Roy \textit{et al.}\cite{9761924}, 2022 & Using generative adverserial networks (GAN) to detect RF fingerprints &  Data generated inhouse & Detection accuracies of more than $99\%$.  No cross-testing.
 \\  
        \hline
        Xing \textit{et al.} \cite{8469002}, 2018& Autoencoders and GAN using multiple stages of the receiver &  Detection accuracy per class&  No cross-testing.
 \\ 
\hline
        Baldini \textit{et al.} \cite{baldini2023convolutional}, 2022 & CNN on data sets preprocessed by substracting the mean and computing the ratio betwen the signal and the mean & Archiutecture tested against two different data sets& Accuracy of more than $90\%$ for high signal to noise ratios. No cross-testing.  \\  
        \hline       
        Jiang \textit{et al.} \cite{9960779}, 2022 & RF fingerprintring indentification using spiking neural network (SNN)  & MATLAB simulation dataset & The proposed solution can reach up to $95\%$ accuracy at high SNR. No cross-testing.  \\  
        \hline
        Huang \textit{et al.} \cite{10568158}, 2024 & Using both GAN and CNN for spoofer deetction & LoRa based data set & Detection accuracy of around $92\%$. No cross-testing. \\  
        \hline
       Ferre \textit{et al.} \cite{morales2020identifying}, 2020 & Used SVM for RF fingerprinting  &  Authors generated raw data for GNSS & Their experimental results show more than $99\%$ classification accuracy. No cross-testing. \\  
        \hline
        Jiang \textit{et al.} \cite{jiang2024radio}, 2024 & Introduced variational autoencoders and LSTM to lear RF fingerprints in GNSS  & Inhouse data set and public data sets & Results of $96\%$ classification. No cross-testing.
 \\  
        \hline
       She \textit{et al.} \cite{10685447}, 2024 & AdaBoost-CNN algorithm for classification of spoofing incidents in UAV tracking& Both simulated and real data  & Up to $95\%$ detection accuracy. No cross-testing.   \\  
        \hline
        Iqbal \textit{et al.} \cite{10495074}, 2024 & Deep-learning based spoofing detection based on post-correlation data & TEXBAT  & Cross-testing accuracies varied between $62\%$ and $78\%$ with best results obtained when training on multiple training datasets. No pre-correlation data used. \\
       \hline
          Marata \textit{et al.} \cite{Marata2025}, 2025 & Four ML algorithms investigated for cross-testing capabilities; best results obtained with Convolutional Neural Networks (CNN), but insufficient cross-testing accuracies &  TEXBAT and OAKBAT  & Cross-testing accuracies varied between $52\%$ and $99.9\%$ with better results obtained when training on  complex spoofing datasets than on simple spoofing datasets  \\ 
       \hline
      {\bf Our current work} & {\bf Meta Learning using both precorelation and postcorelation data} & {\bf TEXBAT and OAKBAT} & {\bf More than $95\%$ detection accuracy, with cross testing.}    \\  
        \hline
    \end{tabular} }
    \label{TableComparison}
\end{table*}

\subsection{Paper Contributions}
In this paper, we consider a GNSS-assisted navigation scenario where a GNSS receiver acquires information from multiple GNSS satellites. At the same time, a software-defined radio generates a spoofing signal with the intention of distorting the navigational data. For this, we develop a framework that enables the GNSS receiver to classify between genuine and spoofing signal. Most importantly, the proposed solution works for different types of spoofying signals. The main contributions are as follows: 
\begin{itemize}
    \item We formulate a spoofing detection problem, which we address using the Model-Agnostic Meta-Learning (MAML) concept. The MAML framework enables machine learning models trained on certain features to adapt to previously unseen features. This is achieved by taking a small subset of a new dataset and using it to refine the model. In our work, we use a model pretrained on different spoofing scenarios and adapt it to novel spoofing scenarios by ensuring the transferability of features. As a result, our proposed model generalizes very well to various spoofing scenarios, outperforming the previously reported studies in the literature on generalizability \cite{10495074,Marata2025}.
    \item We develop our spoofing detection framework to exploit the interdependence of RF fingerprints at both the precorrelation and postcorrelation receiver stages. Specifically, we use the magnitude spectrograms of the in-phase and quadrature (IQ) data for the precorrelation stage, and various postcorrelation features such as code phase. This departs from most existing works, with the exception of our prior work in \cite{9847540}, which typically focus on only one specific stage of the GNSS receiver. Our approach leverages the correlation in the behavior of GNSS signals across different receiver stages, thereby yielding superior performance.  
    \item Finally, we evaluate our proposed solution on TEXBAT and OAKBAT datasets, and compare it with existing solutions. The results showcase superior performance of our proposed solution, both in terms of accuracy and in terms of generalization. For instance, our proposed models successfully detects all the spoofing scenarios in OAKBAT \cite{oakbat} and TEXBAT \cite{texbat} repositories, with an accuracy of more than $99\%$, outperforming the previously reported accuracies in cross-testing \cite{10495074,Marata2025}.
\end{itemize}
\subsection{Organization and Notations}
The remainder of this paper is organized as follows: Section~\ref{ProblemFormulation} presents the system model and the mathematical formulation of the spoofing detection problem, including both feature extraction and data preprocessing stages. Section~\ref{methodology} describes the methodological development of the proposed solution and an overview of the MAML-based approach. Section~\ref{ResultsAndDiscussion} presents the results and discussions. Finally, Section~\ref{conclusion} concludes the paper and outlines directions for future research.

\par 
\textit{\textbf{Notations:}} In this paper, boldface uppercase and lowercase represent matrices and vectors, respectively. The notations, $\circledast$ and $\odot$ represent the convolution and Hadamard product operations, respectively. The notation $\mathcal{N}(\bm{\mu}, \bm{\sigma}^2)$ represents the Gaussian distribution with mean $\bm{\mu}$ and variance $\bm{\sigma}^2$. Fruthermore, we denote We denote sets using calligraphic letters, e.g. $\mathcal{G}$.    
\section{System Model And Problem Formulation}
\label{ProblemFormulation}
\begin{figure}
    \centering    \includegraphics[width=0.99\linewidth]{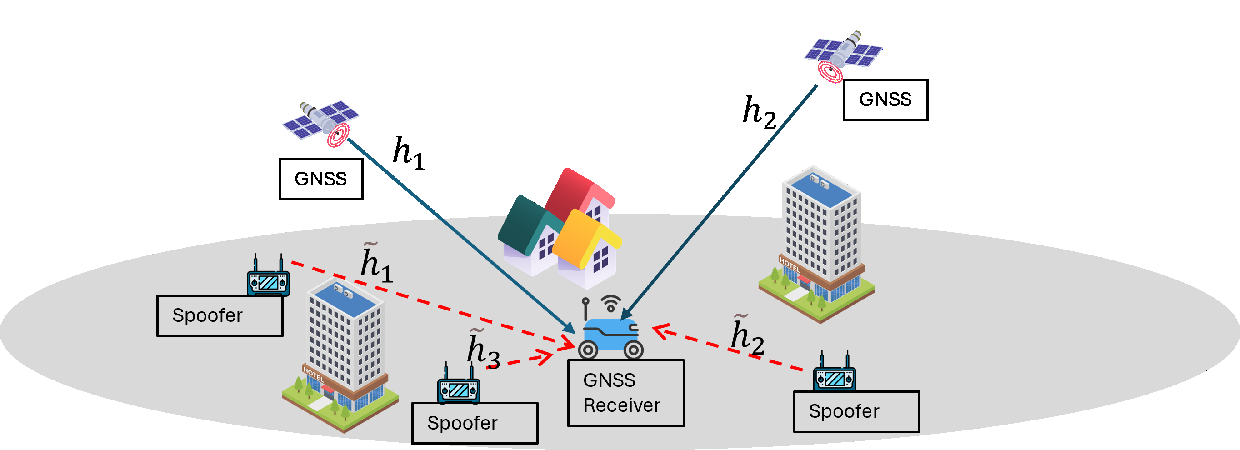}
    \caption{A static spoofing scenario in which multiple spoofers are attempting to deceive a self-driving robot.}
    \label{scenarioSpoofing}
\end{figure}
We consider a static spoofing scenario in which a set $\mathcal{G} = \{1, \cdots, G\}$ of GNSS satellites is visible to a GNSS receiver, as illustrated in Fig.~\ref{scenarioSpoofing}. The receiver also receives spoofing signals generated by software-defined radios (SDRs) with the intention of deceiving it into computing incorrect navigational information. Let the set $\mathcal{N} = \{1, \cdots, N\}$ denote the spoofing signals generated by the SDRs. In the absence of interference (such as spoofers or jammers) and RF imperfections introduced by the transmitting devices, the received GNSS signal can be modeled as the convolution of the transmitted signal and the wireless channel. For a signal $s_g$ from the $g$-th satellite and the wireless channel $h_g$ between the receiver and the $g$-th satellite at time $t$, this becomes \begin{subequations} \label{modelEquation}
\begin{align}
r(t) &= \sum_{g=1}^{G} \sum_{l=1}^{L_g} \alpha_{g,l} \, s_g(t - \tau_{g,l}) + v(t) \label{rfgnssfIRSTEQ} \\
     &= \sum_{g=1}^{G} s_g(t) \circledast h_g(t) + v(t),
     \label{RFGNSSeq2}
\end{align}
\label{ModelOne}
\end{subequations}where $L_g$ in \eqref{rfgnssfIRSTEQ} denotes the number of multipaths in the channel, $\tau_{g,l}$ is the $l$-th path delay for the signal transmitted by $g$ satellite,  $v(t)$ is the Gaussian noise at the receiver. The equation \eqref{RFGNSSeq2} is a more compact form of the received GNSS signal. By considering the presence of the set $\mathcal{N}$ of spoofers, the model \eqref{ModelOne} can be modified and expressed as follows:\begin{subequations} \label{RFModel}
\begin{align}
r(t) &= \sum_{g=1}^{G} \sum_{l=1}^{L_g} \alpha_{g,l} \, s_g(t\!-\!\tau_{g,l}) + 
       \sum_{n=1}^{N} \sum_{k=1}^{K_n} \beta_{n,k} \, \tilde{s}_n(t\!-\!\delta_{n,k}) + v(t)
\label{RFModel_a} \\
     &= \sum_{g=1}^{G} \left[ s_g(t) \circledast h_g(t) \right] + 
       \sum_{n=1}^{N} \left[ \tilde{s}_n(t) \circledast \tilde{h}_n(t) \right] + v(t),
\label{RFModel_b}
\end{align}
\end{subequations}where $k_n$ are the number of multipath components of the channels of each spoofer signal. The notations $\tilde{s}_n(t)$ and $\tilde{h}_n(t)$ are the spoofing signal and their channels, respectively, while \eqref{RFModel_b} expresses the received signal in a more compact form. As mentioned earlier, even devices manufactured to the same specifications tend to differ, as they introduce unique RF fingerprints into the signal. In this scenario of Fig.~\ref{scenarioSpoofing}, where GNSS satellites and SDR devices act as transmitters, the RF fingerprints introduced by their respective hardware will result in distinct impairments. Assuming that the fingerprints coming from the genuine GNSS transmitter are denoted by $f_g(\cdot)$, while for spoofer device are denoted by $\tilde{f}_n(\cdot)$, the equations \eqref{RFModel_a} and  \eqref{RFModel_b} can be modified to become
 \begin{subequations}
\begin{equation}
\begin{split}
r(t) = \sum_{g=1}^{G} \sum_{l=1}^{L_g} \alpha_{g,l} \, f_g\left(s_g(t - \tau_{g,l})\right) 
\\ 
+ \sum_{n=1}^{N} \sum_{k=1}^{K_n} \beta_{n,k} \, \tilde{f}_n\left(\tilde{s}_n(t - \delta_{n,k})\right) + v(t),
\end{split}
\label{FinalModelRF}
\end{equation}

\begin{equation}
r(t) = \sum_{g=1}^{G} \left[ f_g\left(s_g(t)\right) \circledast h_g(t) \right] + 
       \sum_{n=1}^{N} \left[ \tilde{f}_n\left(\tilde{s}_n(t)\right) \circledast \tilde{h}_n(t) \right] + w(t),
\label{eq:rf_convolution_model}
\end{equation}
\end{subequations}respectively. It is important to emphasize that the RF fingerprints can be seen as a function that takes the trasmitted signal as an input. Similar to \eqref{modelEquation}, the resulting signal is a convolution of the channel and the signal distorted by the RF fingerprints. As such, in \eqref{eq:rf_convolution_model}, the problem of spoofing detection is mainly to classify the structure of the received signal based on $\tilde{f}_n$ and ${f_g}$. It is important to note that the resulting signals will exhibit different time-frequency patterns, thereby enabling their classification.  In this work we proposed to solve this problem using learning based approaches, specifically Meta Learning framework. Though we use both pre-correlation and post-correlation data, to perform this, we first preprocess the IQ data into spectrograms to seperate the signal into its time-frequency patterns, as we discuss next.  

\subsection{Feature Extraction}
As discussed earlier, we propose the detection of spoofers by using both the pre-correlation data and the post-correlation data. This is mainly to exploit the correlation that exists between the two stages receiver stages \cite{9847540}. For the pre-correlation phase, a critical step is to extract features from the RF data, which in this case are the IQ samples. This extraction essentially reveals the RF fingerprints that capture the impact of the hardware used to generate the signal, which, in the scenario described under Section~\ref{ProblemFormulation}, could be the GNSS transmitter or the SDR transmitter. Moreover, this can assist in simplifying the classification based on convolution, which can be computationally expensive, by working with products of signals in the frequency domain. We specifically choose channel-independent spectrogram magnitudes for this task\footnote{The phase spectrogram can also be computed as a feature. We have focused on the magnitude spectrogram as the phase spectrogram did not add any advantages to the detection process.}. To achieve this, we compute the Short-Time Fourier Transform (STFT) of the signal. Here, the signal $r(t)$ is divided into $K$ segments, each of $N$ samples \cite{10333639}, thus leading to each point being represented in time and frequency using STFT as 
\begin{equation}
    \bm{R}_{[m,k]} = \sum_{q = 0}^{Q-1}s[q]w[q-mL]\exp{\left(-j\pi \frac{k}{N}q\right)},
\end{equation}
where $k = 1, \cdots, K$ and $m = 1, \cdots, N$ denote the row and column entries of the matrix $\bm{R}\in \mathbb{C}^{N \times K}$ of the STFT and $w[q]$ is the spectral window function, while $L$ is the hop size of the STFT. In frequency domain, the generic RF fingerprint from eq.  \eqref{FinalModelRF} can be represented by $F(\cdot)$, while the channel is represented by $\bm{H}$. As such, the overall impact of the RF fingerprint can be denoted by the Hadamard product of the channel and the RF fingerprints, as follows
\begin{equation}
    \bm{R} = \bm{H}\odot F(\bm{X}),
\end{equation}
where 
\begin{subequations}
\begin{align}
\bm{H}&= 
\begin{bmatrix}
    H_{1,1} & H_{1,2} & \cdots & H_{1,K} \\
    H_{2,1} & H_{2,2} & \cdots & H_{2,K} \\
    \vdots & \vdots & \ddots & \vdots \\
    H_{N,1} & H_{N,2} & \cdots & H_{N,K}
\end{bmatrix} \\\\
F(\bm{X}) &= 
\begin{bmatrix}
    f(X_{1,1}) & f(X_{1,2}) & \cdots & f(X_{1,K}) \\
    f(X_{2,1}) & f(X_{2,2}) & \cdots & f(X_{2,K}) \\
    \vdots & \vdots & \ddots & \vdots \\
    f(X_{N,1}) & f(X_{N,2}) & \cdots & f(X_{N,K})
\end{bmatrix}
\end{align}
\label{EquationsSPECTROGRAMS}
\end{subequations}

\subsection{Dataset preprocessing}
\begin{figure*}[t!] 
    \centering
    \begin{subfigure}{0.3\linewidth}
        \centering        \includegraphics[width=\linewidth]{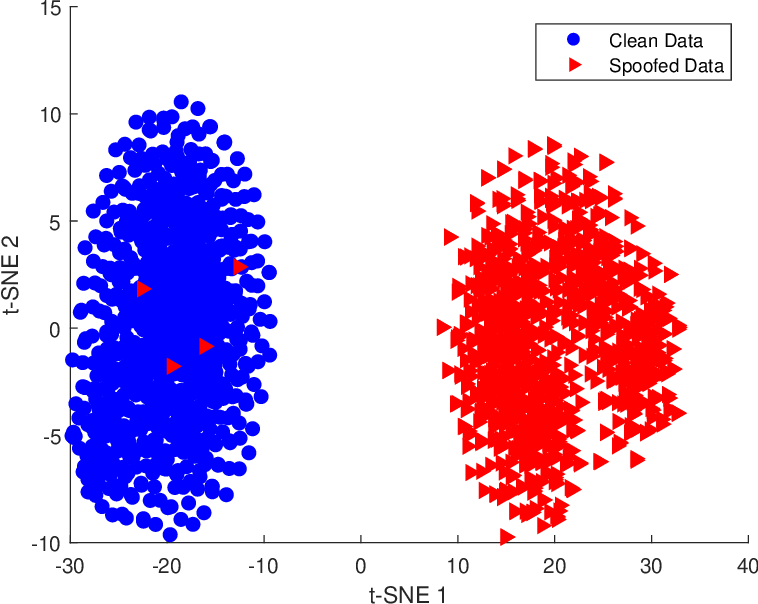}
        \caption{ mag tsne} 
        \label{fig:mag_tsne}
    \end{subfigure}
    \hspace{0.01\linewidth} 
    \begin{subfigure}{0.3\linewidth}
        \centering        \includegraphics[width=\linewidth]{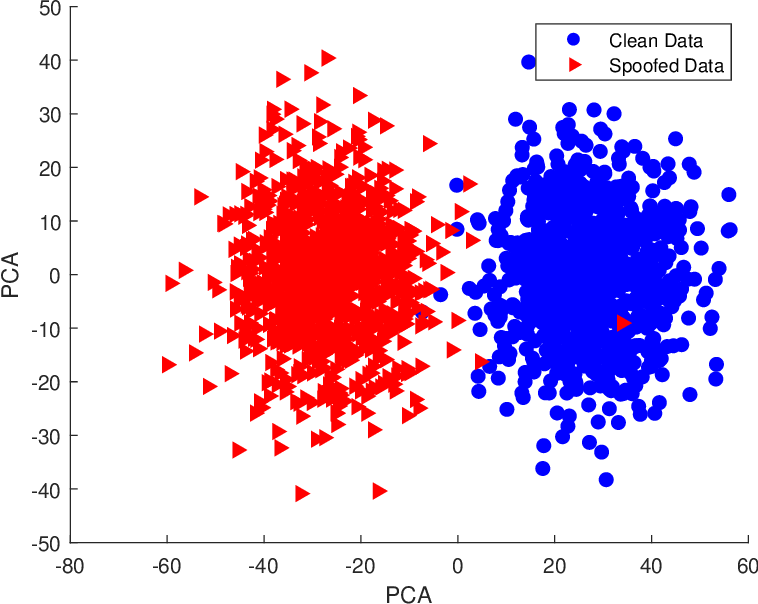}
        \caption{phase pca} 
        \label{fig:phase_pca}
    \end{subfigure}
    \hspace{0.01\linewidth} 
    \begin{subfigure}{0.3\linewidth}
        \centering
        \includegraphics[width=\linewidth]{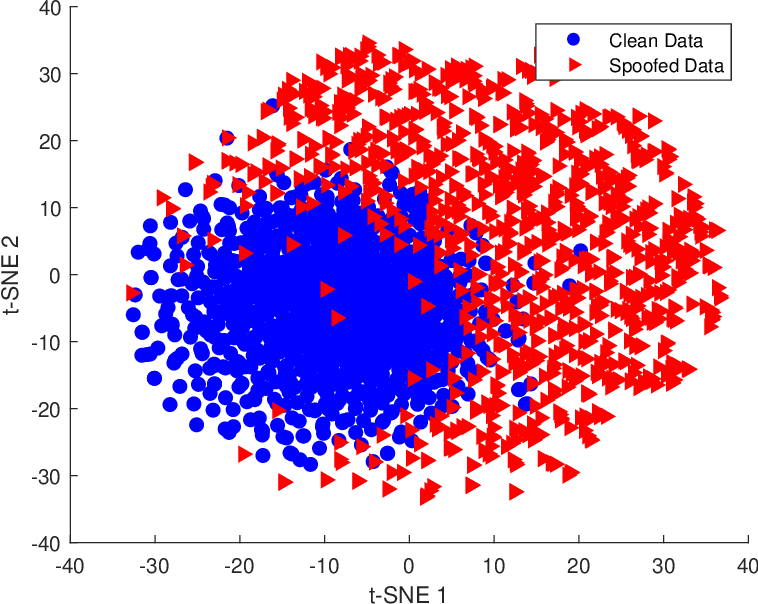}
        \caption{dll tsne} 
        \label{fig:dll_tsne}
    \end{subfigure}
    \vspace{0.5cm} 
    \caption{TEXBAT DS2 dataset, visual representation of feature-based analysis} 
    \label{fig:all_imagesQ}
\end{figure*}

\begin{figure*}[t] 
    \centering
    \begin{subfigure}{0.3\linewidth}
        \centering
        \includegraphics[width=\linewidth]{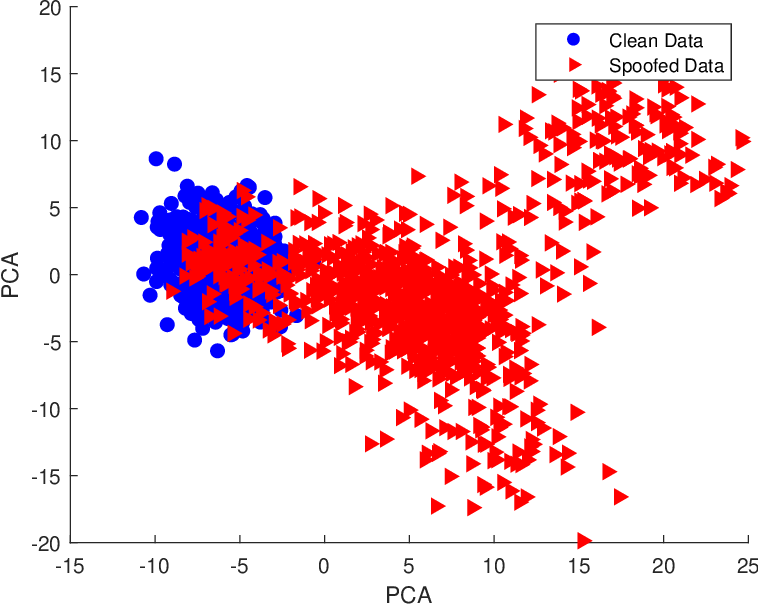}
        \caption{Magnitude pca DS8} 
        \label{fig:mag tsne}
    \end{subfigure}
    \hspace{0.01\linewidth} 
    \begin{subfigure}{0.3\linewidth}
        \centering
        \includegraphics[width=\linewidth]{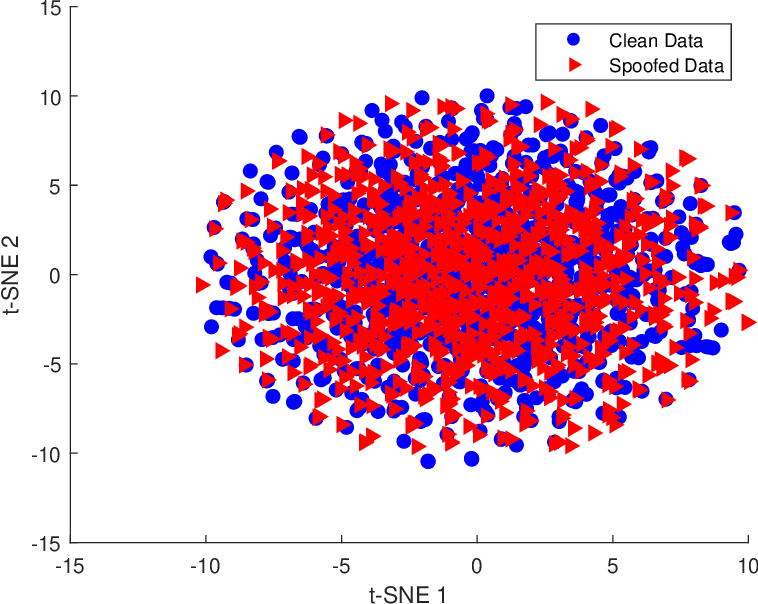}
        \caption{ phase\_pca} 
        \label{fig:phase pca}
    \end{subfigure}
    \hspace{0.01\linewidth} 
    \begin{subfigure}{0.3\linewidth}
        \centering
        \includegraphics[width=\linewidth]{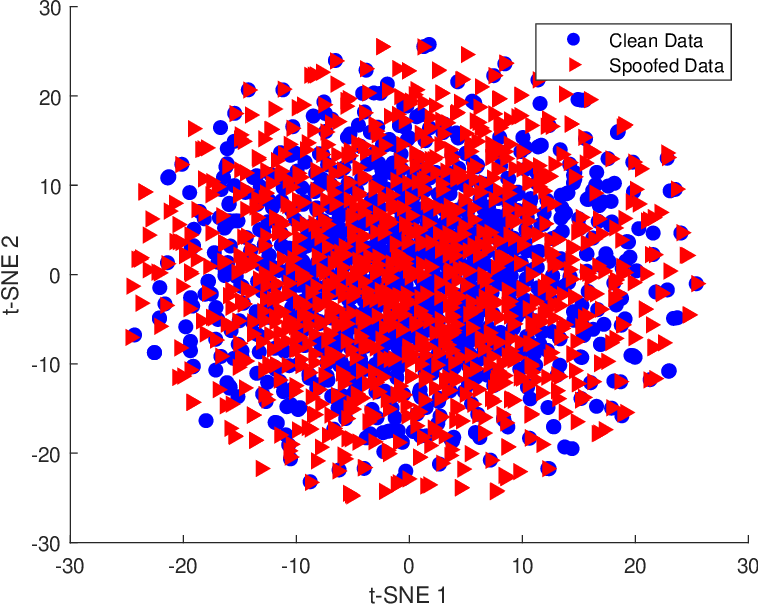}
        \caption{dll\_tsne} 
        \label{fig:dll tsne}
    \end{subfigure}
    \vspace{0.5cm} 
    \caption{TEXBAT DS8 dataset, visual representation of feature-based analysis} 
    \label{fig:all_images}
\end{figure*}

In this work, we specifically use datasets from the TEXBAT and OAKBAT repositories, as mentioned earlier in Section~\ref{STATEOFART}; TEXBAT datasets are referred to as DS or ds, followed by the corresponding datasets index (e.g., DS2, DS8, ...), while OAKBAT datasets as OS or os (e.g., OS2, ...). Full description of each of these datasets and spoofing scenarios corresponding to each dataset index can be found in \cite{texbat,oakbat}  and it is not reproduce here for lack of space. It is to be noted that both TEXBAT and OAKBAT are available in free open access as of August 2025, but OAKBAT access nowadays does require some registration. To align with the generic frameworks of \eqref{EquationsSPECTROGRAMS}, we preprocessed the data by capturing $60$ seconds worth of data for each category, spoofed and clean. We thereafter, segment each data set into $4$ ms sequences. This results in $15000$ spectrograms samples per category, making a total of $30000$ spectrograms. Similarly, we process the post-correlation data using the FGI-GSRx software receiver \cite{liaquat2024enhanced}, applying the same segmentation duration, in order to extract also the postcorrelation features (i.e., features after the tracking stage). As an example, we present visual representations of the different datasets using Principal Component Analysis (PCA) and t-SNE, shown in Fig.~\ref{fig:all_images} and Fig.~\ref{fig:all_imagesQ} for two example datasets (TEXBAT DS2 and DS8). As can be noted from the Figs.~\ref{fig:all_images} and ~\ref{fig:all_imagesQ}, there can be a clearcut boundary between features in some data sets such as for DS2. However, there are some spoofing cases such as the DS8, where the boundary is not obvious as can be noted for the Fig. ~\ref{fig:all_imagesQ}. As such, most ML based solutions cannot be trained on one data set and generalized to completely unseen data, as also our prior work in \cite{Marata2025} has shown. This has been the case with TEXBAT repository data, where simple spoofing scenarios do not generalize well to other types of spoofing. To address this, we solve the problem using the Meta learning model agnostic learning, which we present next. 
\section{Methodology and Proposed Solutions}
\label{methodology}
\begin{figure}[t!]
    \centering    \includegraphics[width=0.99\linewidth]{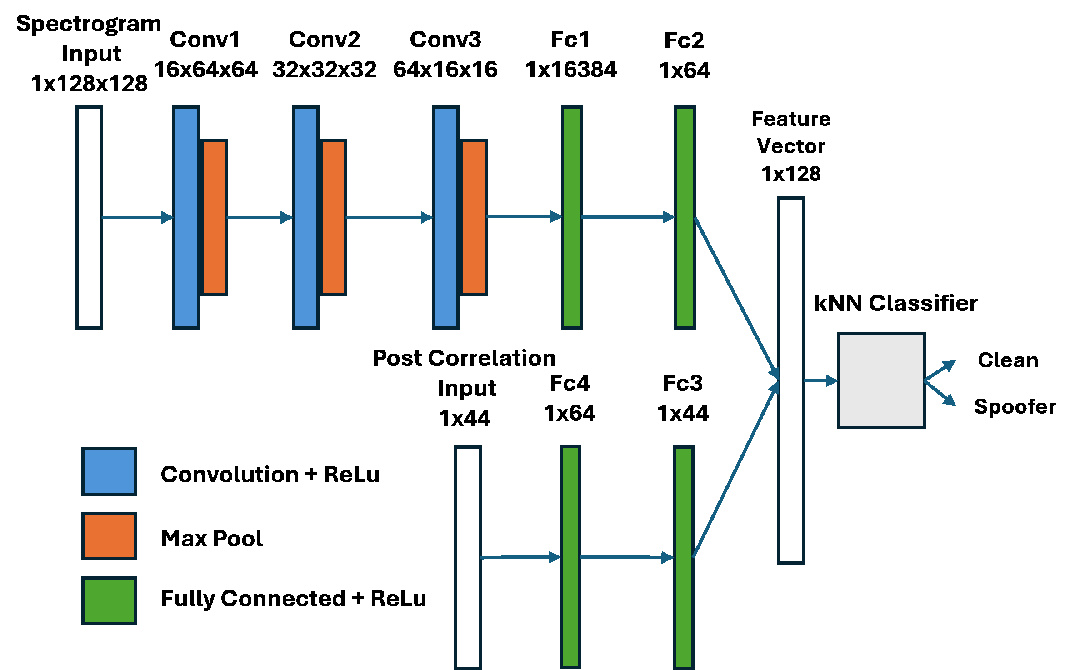}
    \caption{Architecture of dual-branch CNN embedding model (spectrogram + post correlation) with k-NN classifier.}    \label{modelArch}
\end{figure}

As mentioned earlier, most learning algorithms, such as neural networks require large amounts of labeled data. Unfortunately, such data can often be expensive to obtain or may simply be unavailable. For example, spoofers are constantly evolving, and each new variant may exhibit entirely unseen characteristics to GNSS receivers, thereby presenting novel challenges. As a result, even AI-powered GNSS receivers may struggle to adapt to emerging threats, posing significant risks to GNSS users.
To address this limitation, we enhance the Convolutional Neiural Networks (CNNs), previously shown to have the best results among investigated ML algorithms \cite{Marata2025} with the ability to efficiently adapt to different types of spoofers, including those they have not encountered before. We achieve this by leveraging the Model-Agnostic Meta-Learning  concept, capitalizing on its ability to \textit{learn how to learn}. The basic architecture of the model is presented in Fig.~\ref{modelArch}. To make this paper self-contained, we begin by providing an overview of the MAML algorithm.
\subsection{An overview of MAML}
The central idea behind MAML is to increase the model’s sensitivity to subtle variations in signal structure, thus allowing it to quickly adapt to new spoofing patterns \cite{pmlr-v70-finn17a}. To provide context, it is important to note that each conventional learning problem such as CNN can be defined as
\begin{equation}
\mathcal{T}_i = { \mathcal{L}(\mathcal{D}_i, \bm{\theta}_i) },
\label{basicTask}
\end{equation}
where $\bm{\theta}_i$ is the vector with the model paramters, $\mathcal{D}_i$ represents the $i$-th dataset comprising input-output pairs, i.e., $\mathcal{D}_i = \{(x_j, y_j)\}_{j=1}^{J}$, sampled from some joint distribution $p(x, y) = p(y|x)p(x)$. In the context of spoofing detection, this formulation allows the model to learn the parameters (e.g., CNN weights) that map specific spoofing characteristics from the data $x$ to their corresponding labels contained in $y$. As such, the trained model should be able to recognize  and classify most of the testing datasets that have some similarities with the data used during training. Unfortunately, if a new dataset with completely different and previously unseen features is presented to the algorithm, the framework can fail to classify the signal accordingly. The MAML framework overcomes this challenge by enabling the model to generalize to completely unseen data, and thus to achieve good cross-testing or generalization capabilities, as we discuss next.

\par In contrast to the formulation in eq. \eqref{basicTask}, the MAML framework generalizes the models by training over multiple tasks $\{\mathcal{T}_i\}_{i=1}^{T}$, where each task is associated with a dataset $\mathcal{D}_i$, for $i \in \{1, 2, \cdots, T\}$. As such, MAML is similar to multi-task learning \cite{zhang2021survey,9084370}, except that the MAML explicitly optimizes the parameters for transferability, enabling efficient fine-tuning on new data. Let us denote the completely unseen spoofing task as $\mathcal{D}_j$. In our case, $\mathcal{D}_j$ is composed of the RF fingerprints $\bm{R}$ (based on the spectrogram magnitude of the precorrelation data)  and the postcorrelation data, such as the discrimination locked loop (dllDiscr), code phase, Doppler, frequency locked loop (fllLock), and phase locked loop (pllLock) outputs. To implement the MAML framework, we split $\mathcal{D}_j$ into two subsets: the support set $\mathcal{D}_j^{\text{support}}$ and the query set $\mathcal{D}_j^{\text{query}}$. The support set is used to quickly adapt the model parameters, while the query set is used to evaluate the adapted model’s performance \cite{10413635}. The model's adapted parameters $\bm{\theta}_j'$ can be computed from the paramters of the model develped at meta training as follows 
\begin{equation}
\bm{\theta}_j' = \bm{\theta} - \alpha \nabla_{\bm{\theta}} \mathcal{L}(f_{\theta_j}),
\label{gradStep}
\end{equation}
where $\alpha$ is the learning rate of the inner-loop update and $f_{\theta_j}$ is the model with the parameters. Note here that the cost function in \eqref{gradStep} is computed using $\mathcal{D}_j^{\text{support}}$. For our work, we use a model based on lightweight dual-branch CNN as a feature encoder and trained within a prototypical meta-learning framework. As such, we target to minimize the sum of the losses across all sampled tasks using the adapted weights as follows\cite{10413635}\begin{equation}    
\underset{\bm{\theta}}{\mathrm{min}}\sum\limits_{\mathcal{T}_i}\mathcal{L}_{\mathcal{T}_j}(\bm{\theta}_j', \mathcal{D}_j^{\text{query}}),
\end{equation}
where $\theta_j'$ is defined in \eqref{gradStep}. For this problem, we adopt a prototypical learning strategy, where task-specific classification is performed using distance-based inference in a learned embedding space. We also introduce sparsity in the weights of the fully connected layer 
This is mainly to introduce regularization into the inner loops, which can improve generalization in few-shot spoofing-detection scenarios \cite{10040221,ADMMMETA}.  To solve this problem, we redefine these weight otpimization as $\ell_1$ minimization framework and solve it via Alternating Direction Method of Multiplers (ADMM) as follows\cite{boyd2011distributed}
\begin{equation}
    \operatorname*{minimize}_{\bm{\theta}} \quad  \sum_{\mathcal{T}_i} \mathcal{L}_{\mathcal{T}_i}(\bm{\theta}) + \lambda \lVert \bm{\theta} \rVert_1,
    \label{ADMM_MAMLL1}
\end{equation}
where $\lVert \bm{z} \rVert_1$ is the sparsity promoting regularization term, while $\lambda$ is the penalization term. This in the ADMM framework becomes
\begin{subequations}\label{eq:ADMM_MAML}
\begin{align}
\operatorname*{minimize}_{\bm{\theta}} & \quad \sum_{\mathcal{T}_j} \mathcal{L}_{\mathcal{T}_j}(\bm{\theta})) + \lambda \lVert \bm{z} \rVert_1, \label{ADMM_MAML12} \\
\text{subject to} & \quad \bm{\theta} = \bm{z}, \label{ADMM_MAML2}
\end{align}
\end{subequations} where $\bm{z}$ is the splitting/decoupling variable. To this end, the weights optimization problem $\bm{\theta}$ can be written as an argumeted Langrangian as follows \cite{boyd2011distributed}
 \begin{equation}
     \mathcal{L}_{\rho}(\bm{\theta}, \bm{z}, \bm{u} ) = \sum_{\mathcal{T}_i} \mathcal{L}_{\mathcal{T}_i}(\bm{\theta}) + \lambda \lVert \bm{z} \rVert_1 + \frac{\rho}{2}\lVert \bm{\theta}- \bm{z}+ \bm{u}\rVert_2^2 + \frac{\rho}{2}\lVert \bm{u}\rVert_2^2,
     \label{ADMM_MAML_L3}
 \end{equation}
where, $\bm{u}$ is the scaled dual variable. This formulations allow for the Meta Loss being minimized by alternatingly updating the varibles, \emph{i.e.}, one variable is updated one at a time, while others are held constant. In this case, for the variable $\bm{\theta}$, the Lagrangian in \eqref{ADMM_MAML_L3} is differentiated with respect to $\bm{\theta}$ and set to zero. Assuming a $t$-th iteration, this yields
\begin{equation}
    \bm{\theta}^{(t+1)} = \operatorname*{min}_{\bm{\theta}} \sum_{\mathcal{T}_i} \mathcal{L}_{\mathcal{T}_i}(\bm{\theta}) + \frac{\rho}{2}\lVert \bm{\theta} -\bm{z}^{(t)} + \bm{u}^{(t)}\rVert_2^2,
\end{equation}
which is solved using the gradient step \eqref{gradStep}. In a similar manner, $\bm{z}$ is updated by holding the other variables constant and finding the derivative of \eqref{ADMM_MAML_L3}, which is then set to zero. As such, at the $(t +1)$- th update, this becomes 
\begin{equation}
    \bm{z}^{(t+1)} = \operatorname*{min}_{\bm{\theta}}  \sum_{\mathcal{T}_i} \mathcal{L}_{\mathcal{T}_i}(\bm{\theta}) + \frac{\rho}{2}\lVert \bm{\theta}_j^{(t+1)} -\bm{z} + \bm{u}^{(t)}\rVert_2^2.\label{proximalOperator}
\end{equation}
From \eqref{proximalOperator}, it can be noted that the derivative results in a proximal operator, which can be written in compact form using a soft thresholding step\cite{boyd2011distributed}, which is given by ${\bm{z}^{(t+1)}= \text{soft}
(\bm{z}^{(t)} + \bm{u}^{(t)},\frac{\lambda}{\rho})}$. Lastly the dual variable is updated by
\begin{equation}
    \bm{u}^{(t+1)} = \bm{u}^{t} + \bm{\theta}_j^{(t+1)}- \bm{z}^{(t+1)}.
\end{equation}
A summary of the proposed framework is provided in Algorithm~\ref{PseudoADMMMAML}.

\begin{algorithm}
  \KwIn{$\{\mathcal{T}_i\}_{i=1}^{T}$,  $\alpha$, $\beta$, $\rho$, $\lambda$, $E$, $K$}  
  \KwData{Initialize model parameters $\bm{\theta}^{(0)}$, dual variable $u^{(0)} = 0$, auxiliary variable $z^{(0)} = \bm{\theta}^{(0)}$}

  \For{epoch $e = 1$ \KwTo $E$}{
    Sample batch of tasks $\{\mathcal{T}_i\}_{i=1}^{B}$\;

    \For{each task $\mathcal{T}_i$}{
      Split into $\mathcal{D}_i^{\text{support}}$, $\mathcal{D}_i^{\text{query}}$\;

      \tcp{Inner Loop: Task adaptation}
      Initialize $\bm{\theta}_i^{(0)} \leftarrow \bm{\theta}$\;

      \For{step $k = 1$ \KwTo $K$}{
        Compute loss $\mathcal{L}_{\text{support}}(\bm{\theta}_i^{(k-1)}; \mathcal{D}_i^{\text{support}})$\;
        Update: $\bm{\theta}_i^{(k)} \leftarrow \bm{\theta}_i^{(k-1)} - \alpha \nabla_{\bm{\theta}} \mathcal{L}_{\text{support}}$\;
      }

      \tcp{Evaluate on query set}
      Compute $\mathcal{L}_{\text{query}}(\bm{\theta}_i^{(K)}; \mathcal{D}_i^{\text{query}})$\;
    }

    \tcp{Meta-update (Outer Loop)}
    Compute meta-gradient: $\nabla_{\bm{\theta}} \sum_i \mathcal{L}_{\text{query}}(\bm{\theta}_i^{(K)})$\;
    Update: $\bm{\theta} \leftarrow \bm{\theta} - \beta \nabla_{\bm{\theta}} \sum_i \mathcal{L}_{\text{query}}(\bm{\theta}_i^{(K)})$\;

    \tcp{ADMM Regularization}
    $\bm{z} \leftarrow \text{prox}_{\lambda/\rho}(\bm{\theta} + \bm{u})$ \tcp*{e.g., soft-thresholding}\;
    $\bm{u} \leftarrow \bm{u} + (\bm{\theta} - \bm{z})$\;
    $\bm{\theta} \leftarrow \bm{z} - \bm{u}$ \tcp*{Update model with regularized weights}
  }
  \KwOut{Meta-trained model parameters $\bm{\theta}$}
  \caption{MAML with ADMM-based Regularization}
  \label{PseudoADMMMAML}
\end{algorithm}

\begin{table}
    \centering
  \caption{Summary of the simulation parameters}
 \begin{tabular}{  m{14em}  m{10em} } 
\hline
\textbf{Parameter} & \textbf{Value}\\ 
\hline
Inner Learning Rate ($\alpha$) & 0.01 \\ 
Outer Learning Rate($\beta$) & 0.001 \\ 
Learning Epochs ($E$) & 8 \\
Querry set size ($K$) & 50 \\
Inner Optimizer  & SDG+ ADMM   \\ 
Outer Optimizer & Adam\\
\hline
\end{tabular}
\label{simulationPar}
\end{table}

\section{Results and Discussions}
\label{ResultsAndDiscussion}
In this section, we present evaluation results of our meta-learning framework. We have adopted to use different data sets from publicly available TEXBAT repositiories and OAKBAT repositories,namely $ds_i$, $i\in {1,2, \cdots, 8}$ and $os_i$, $i\in {2,3,6}$. As such, our experiment considers different spoofing scenarios, ranging from simple to complex spoofing scenarions. The simulation parameters are summarized in Table~\ref{simulationPar}. For the performance analysis, we used the classification accuracy and loss convergence as metrics. The classification performance is illustrated using the confusion matrix. The average value along the diagonal of the confusion matrix reflects the model's accuracy in correctly classifying spoofed signals as spoofed and genuine signals as genuine. In addition to accuracy, we analyze the convergence behavior of the learning algorithms through loss curves. 
\begin{figure}[t!]
    \centering
    \includegraphics[width=0.85\linewidth]{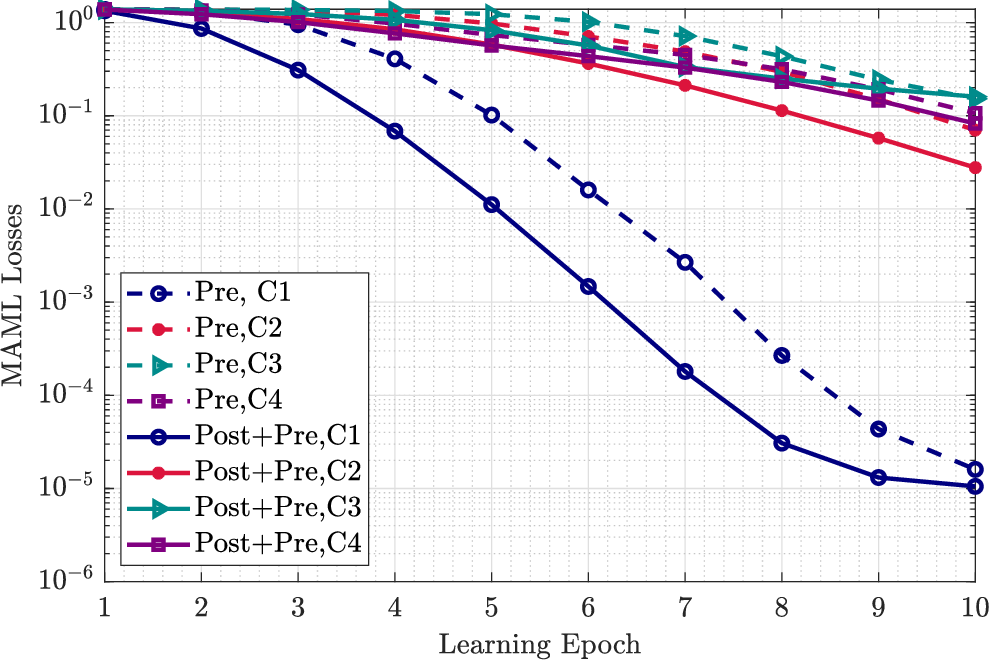}
    \caption{Meta losses as a function of signals at different stages of the receiver, where $C_{i}$ coresponds to the data combinmations; Pre stands for using only precorrelation spectrogram feature and Post+Pre means using both pre- and postcorrelation features.}
    \label{FigPrePost}
\end{figure}
\subsection{On the Impact of Features on learning losses}
\begin{figure}[t!]
    \centering
    \includegraphics[width=0.85\linewidth]{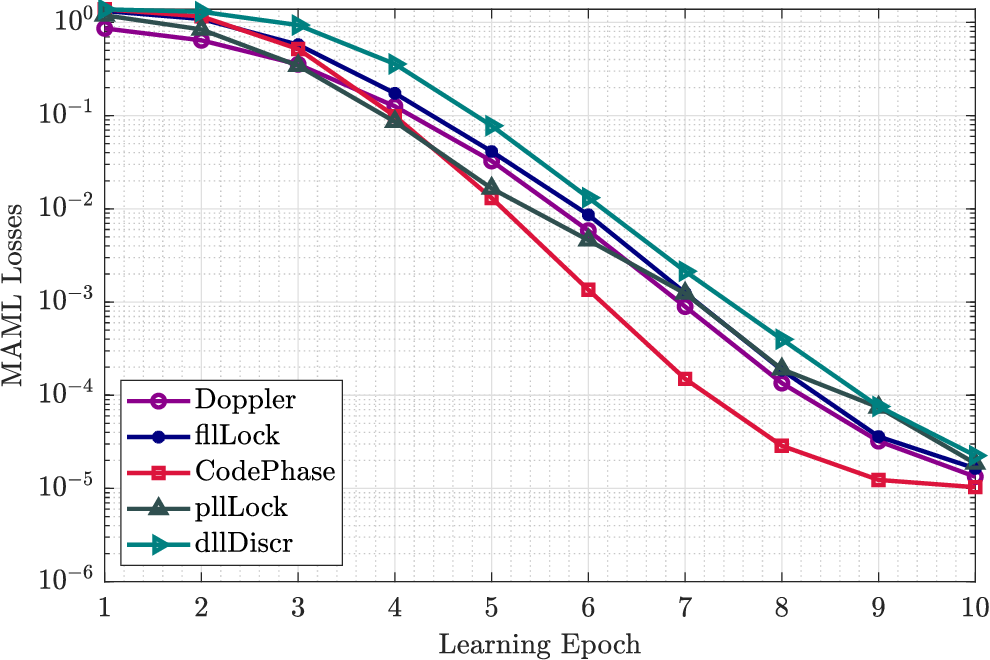}
    \caption{Meta losses as a function of the different postcorrelation features in a GNSS receiver}
    \label{figPostParameters}
\end{figure}

As discussed earlier, to fully benefit from the RF fingerprints, it is necessary to exploit the correlation between the different stages of the signal. In Fig.\ref{FigPrePost}, we present the results of the meta losses when using only precorrelation data and when using both precorrelation and postcorrelation data  (code phase) for different datasets. In the figure $C_i$, $i= 1, 2,3,4$ denotes the combinations of data sets. Here $C_1$ is the learning based on DS2 and DS3, $C_2$ is DS4 and DS7, $C_3$  is DS7 and DS8, while $C_4$ is DS3 and DS8. As it can be seen from Fig. \ref{FigPrePost}, the MAML performance differs across datasets. This essentially shows that ML algorithms are highly sensitive to different forms of spoofers, thus requiring an innovative approach to generalize performance across datasets. Furthermore, it can be observed that when the proposed solution is trained only on precorrelation data, it achieves lower performance compared to when both precorrelation and postcorrelation data are considered. For instance, at the 6th learning epoch, the MAML trained on precorrelation spectrograms achieved a loss of approximately 0.01. However, when some postcorrelation data (in this case, the code phase) is included, the loss is approximately 0.001, which is a tenfold improvement. These results confirm the need to exploit the correlation between different GNSS receiver stages, as this can enhance its capabilities. However, some postcorrelation parameters yield better performance than others, as we show in Fig~\ref{figPostParameters}.

\begin{figure}[t!]
    \centering    \includegraphics[width=0.85\linewidth]{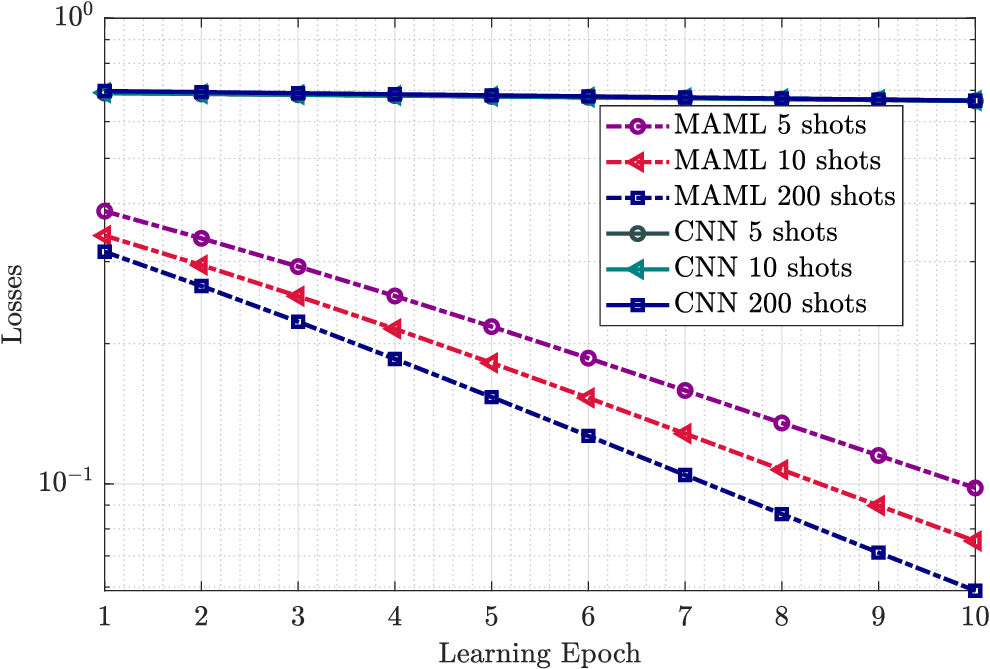}
    \caption{Losses as a function of learning epochs and the number of training shots for CNN and proposed MAML}
    \label{comparsionMAMLVsCNN}
\end{figure}

Fig~\ref{figPostParameters} shows the results of the proposed MAMLframework when using both the precorelation spectrograms and different postcorelation features. As it can be noted in Fig~\ref{figPostParameters}, the general trend shows that the MAML losses decreases with every epoch, which shows that indeed the model is learning the parameters, as shown by losses of approximately $10^{-5}$ after the $10$-th epoch. Inspite of this, the performance still varies across different paramters as shown, which is consistent with works \cite{9844986}. For instance, the learning performance improves when the codephase is used as a feature as compared to when using the  delay lock loop (DLL) discriminator. As such, for the results of the next section, we exploit the spectrograms with codephase data.

Fig.~\ref{comparsionMAMLVsCNN} shows the performance as a function of the number of data points used during the training phase. A general trend shows that the losses reduce with the increase in the number of training samples and the number of learning steps, mainly due to better convergence. Moreover, this is due to better feature exploitaion by learning algroithms. Interestingly, the proposed MAML framework outperforms the CNN framework, even with very few data points for training. For instance, the proposed MAML achieves a loss of less than $0.1$ when using just $5$ labeled samples per class for training, while the CNN based approach achieves more than $0.6$ in losses for $200$ learning epochs or samples. Essentially, this shows that the the MAML proposed framework is highly efficient in utilizing the features, which is consistent with results in \cite{10948463}. Moreover,  one of the main advantages of the MAML is  its ability to optimize the weights such that they can be adapted to unseen data and this is what makes it feasible  to generalize the detection to diverse spoofing types, which makes our proposal of practical interest.
\subsection{Classification Accuracies}
\begin{figure*}[t!]
    \centering

    \begin{minipage}{\textwidth}
        \centering
        \begin{subfigure}{0.3\textwidth}
            \includegraphics[width=\linewidth]{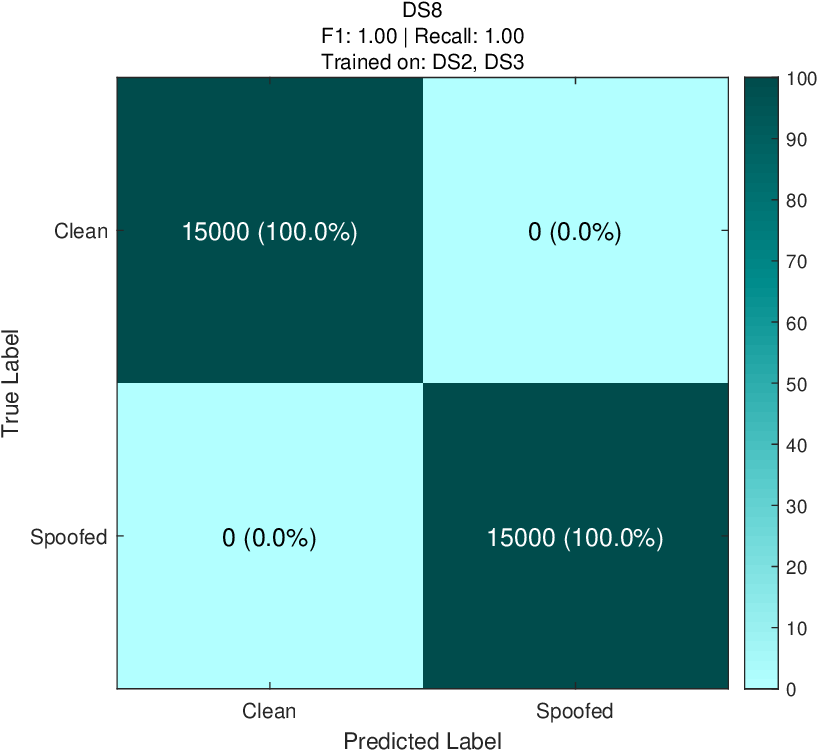}
            \label{fig:confmat_ds8}
        \end{subfigure}
        \hspace{0.03\textwidth}
        \begin{subfigure}{0.3\textwidth}
            \includegraphics[width=\linewidth]{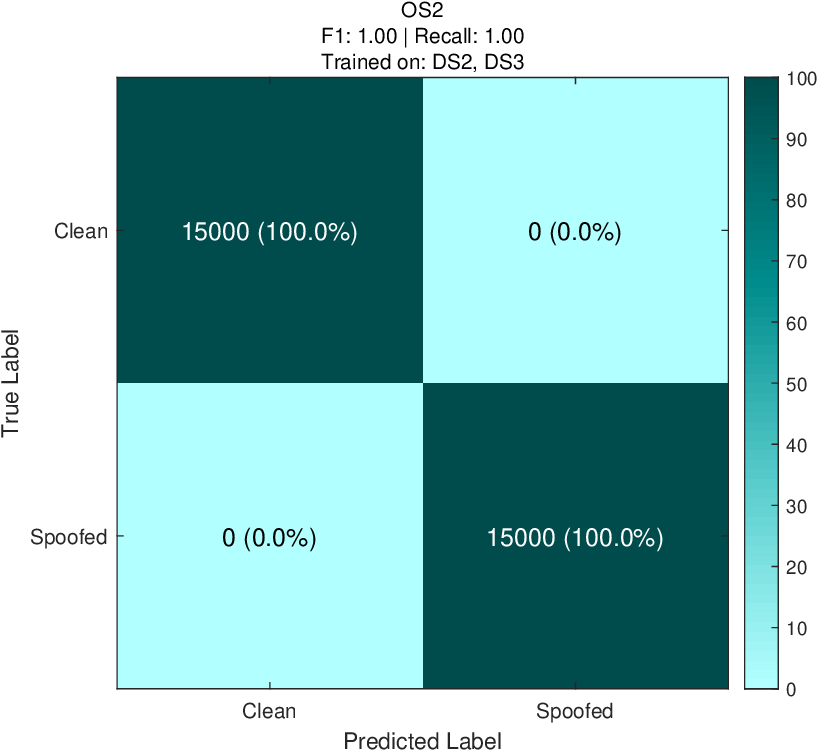}
            \label{fig:confmat_os2}
        \end{subfigure}
        \hspace{0.03\textwidth}
        \begin{subfigure}{0.3\textwidth}
            \includegraphics[width=\linewidth]{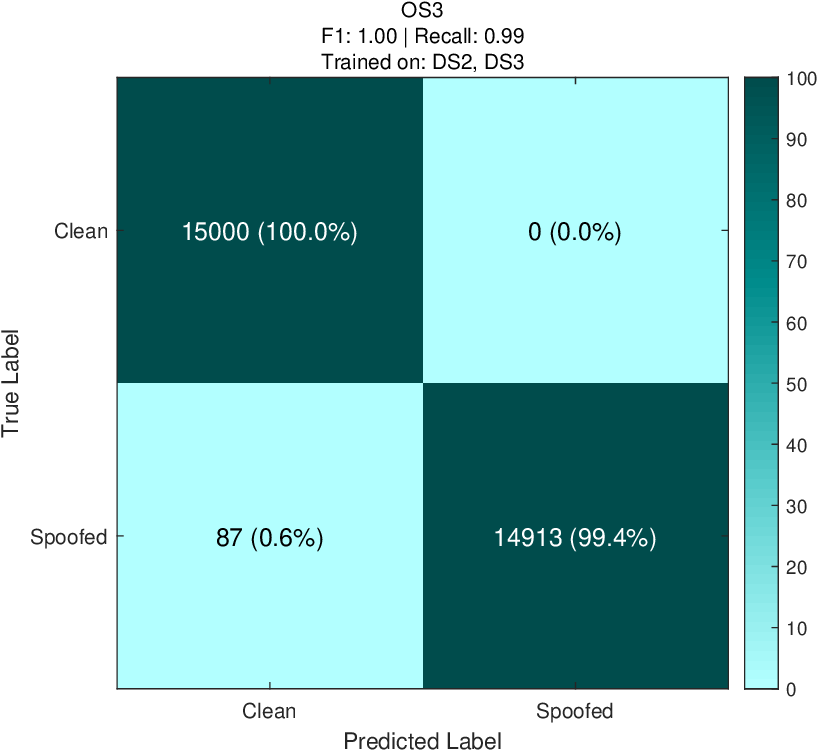}
            \label{fig:confmat_os3}
        \end{subfigure}
    \end{minipage}

    \vspace{0.5em} 

    \begin{minipage}{\textwidth}
        \centering
        \begin{subfigure}{0.3\textwidth}
            \includegraphics[width=\linewidth]{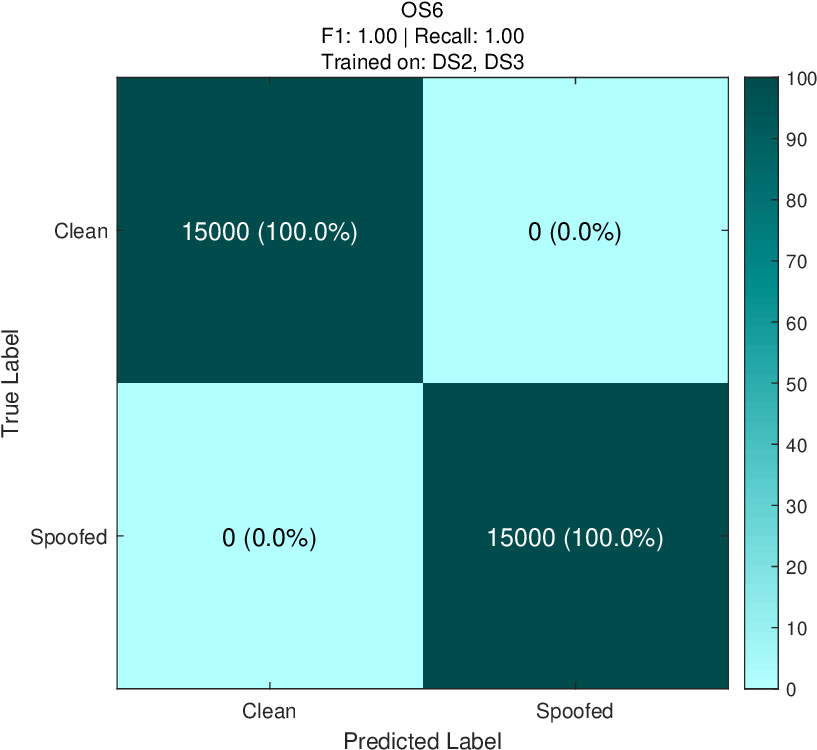}
            \label{fig:confmat_os6}
        \end{subfigure}
        \hspace{0.03\textwidth}
        \begin{subfigure}{0.3\textwidth}
            \includegraphics[width=\linewidth]{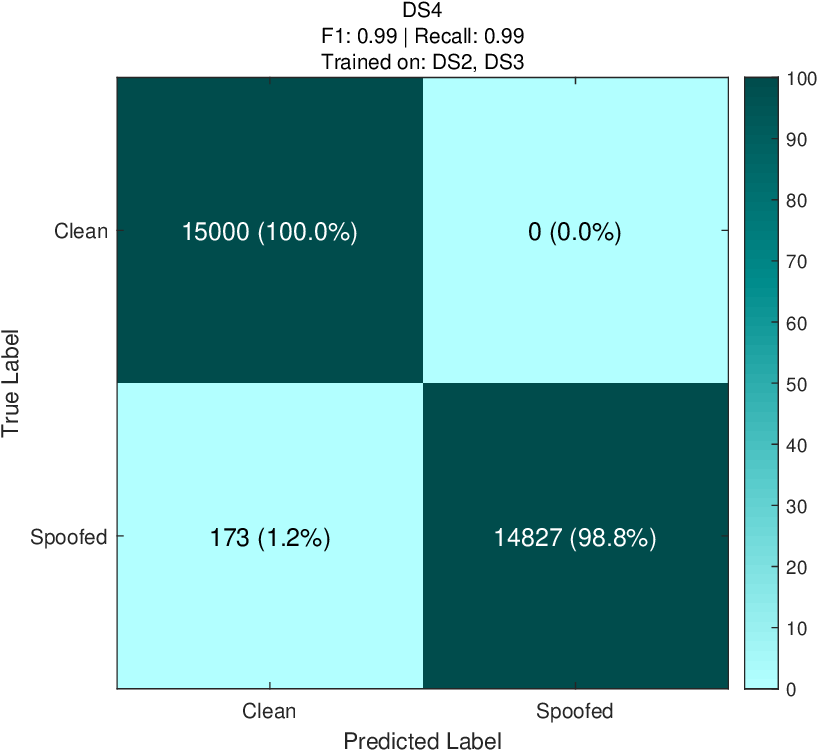}
            \label{fig:confmat_ds4}
        \end{subfigure}
        \hspace{0.03\textwidth}
        \begin{subfigure}{0.3\textwidth}
            \includegraphics[width=\linewidth]{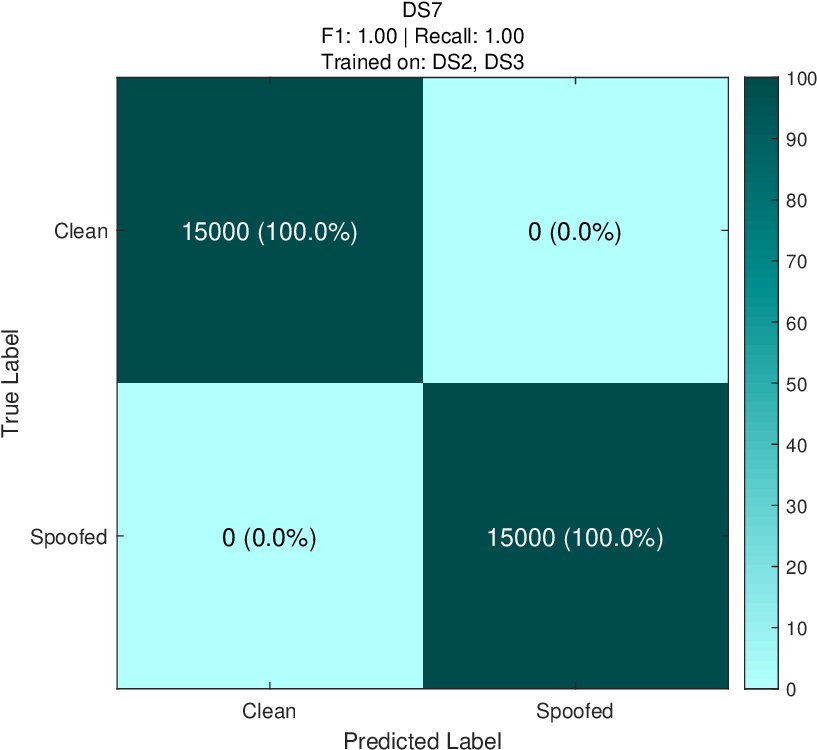}
            \label{fig:confmat_ds7}
        \end{subfigure}
    \end{minipage}

    \caption{Confusion matrices for different spoofing types for the meta-learner trained on TEXBAT DS2 and DS3. The cross-testing datasets is shown on top of each figure.}
    \label{confusion1}
\end{figure*}

\begin{figure*}[t!]
    \centering

    \begin{minipage}{\textwidth}
        \centering
        \begin{subfigure}{0.3\textwidth}
            \includegraphics[width=\linewidth]{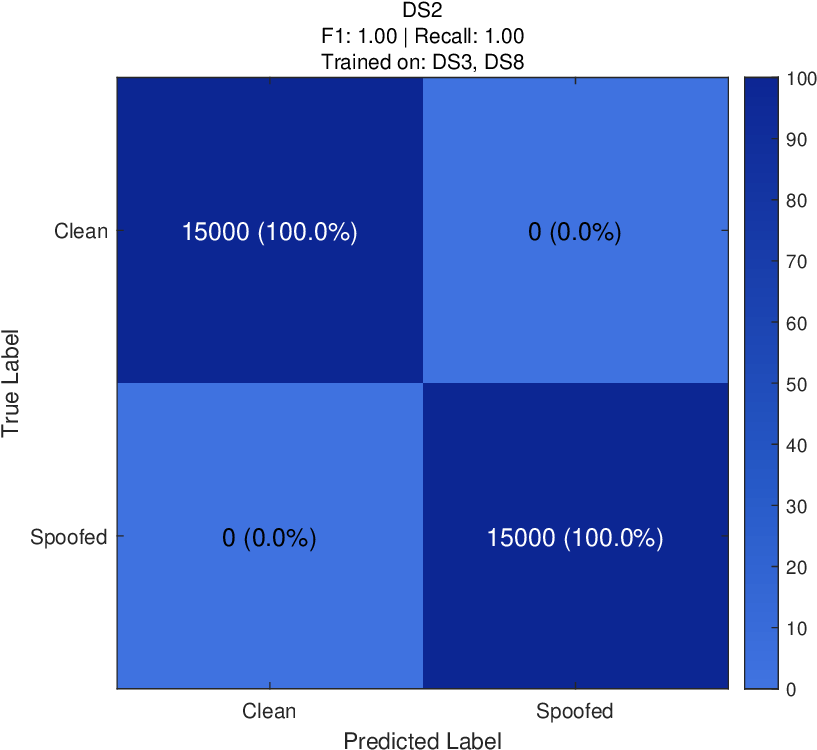}
            \label{fig:confmat_ds8_ds2}
        \end{subfigure}
        \hspace{0.03\textwidth}
        \begin{subfigure}{0.3\textwidth}
            \includegraphics[width=\linewidth]{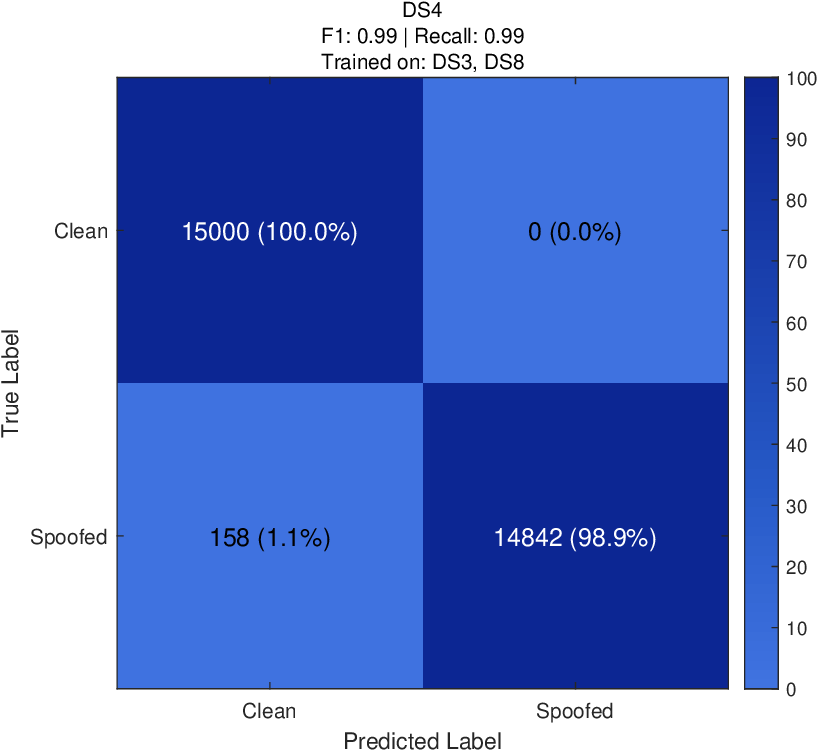}
            \label{fig:confmat_ds8_ds4}
        \end{subfigure}
        \hspace{0.03\textwidth}
        \begin{subfigure}{0.3\textwidth}
            \includegraphics[width=\linewidth]{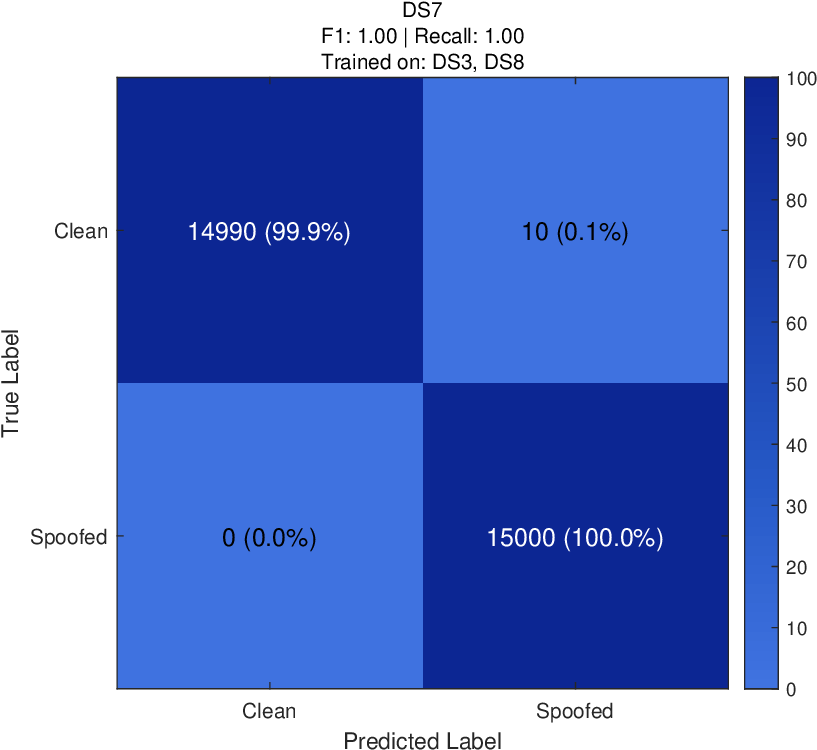}
            \label{fig:confmat_ds8_ds7}
        \end{subfigure}
    \end{minipage}

    \vspace{0.5em} 

    \begin{minipage}{\textwidth}
        \centering
        \begin{subfigure}{0.3\textwidth}
            \includegraphics[width=\linewidth]{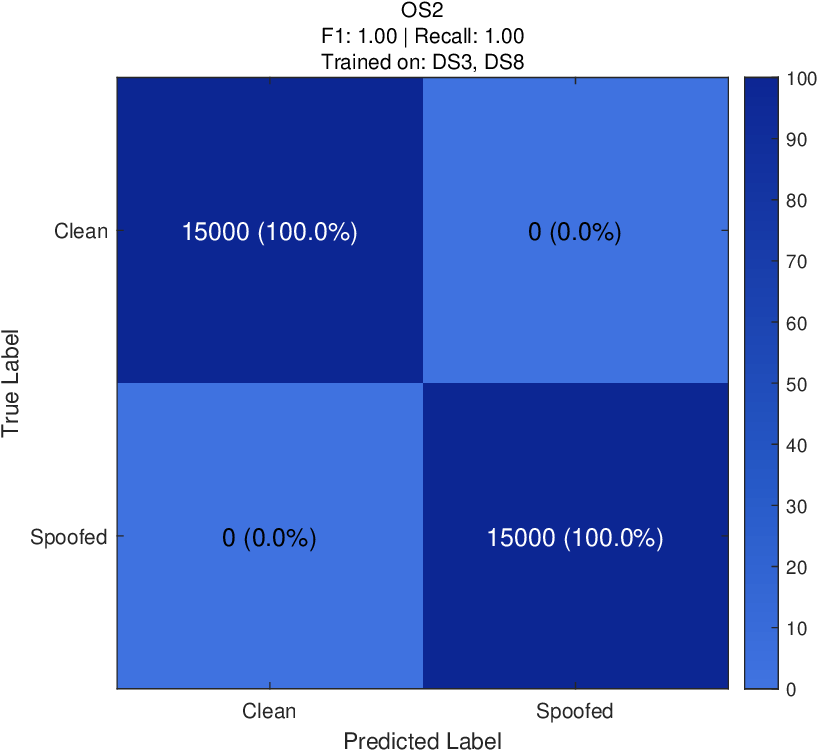}
            \label{fig:confmat_ds8_os2}
        \end{subfigure}
        \hspace{0.03\textwidth}
        \begin{subfigure}{0.3\textwidth}
            \includegraphics[width=\linewidth]{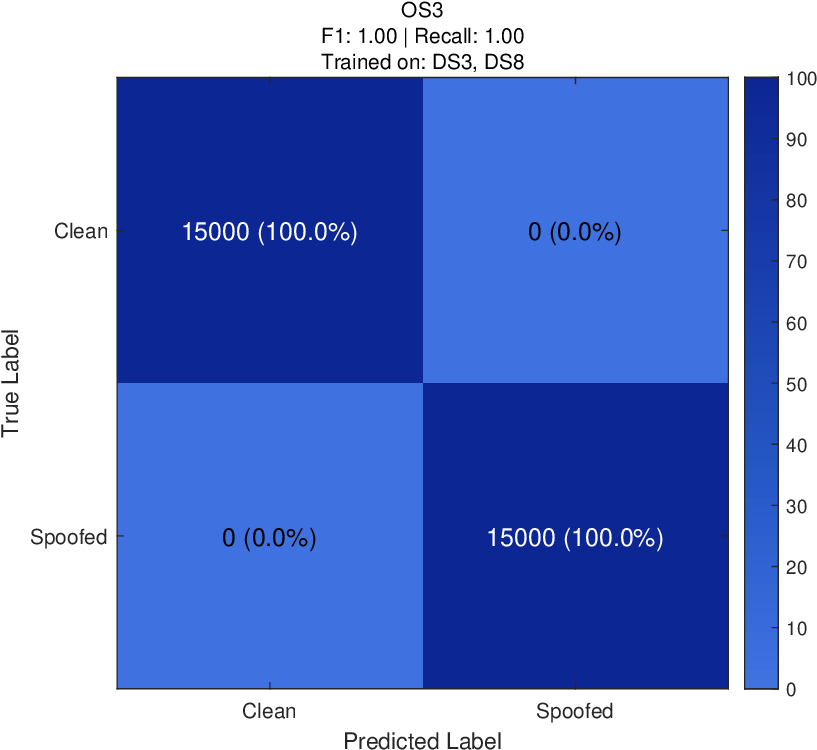}
            \label{fig:confmat_ds8_os3}
        \end{subfigure}
        \hspace{0.03\textwidth}
        \begin{subfigure}{0.3\textwidth}
            \includegraphics[width=\linewidth]{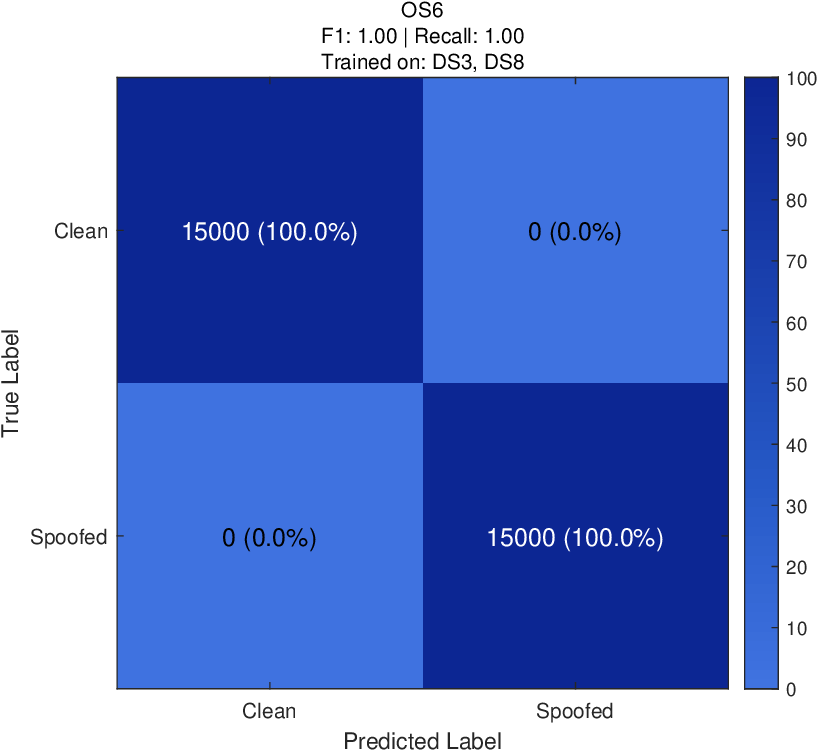}
          \label{fig:confmat_ds8_os6}
        \end{subfigure}
    \end{minipage}

    \caption{Confusion matrices for different spoofing types for the meta-learner trained on TEXBAT DS2 and DS8. The cross-testing datasets is shown on top of each figure.}
    \label{confusion2}
\end{figure*}

Figs.~\ref{confusion1} and \ref{confusion2} present the classification results using the confusion matrices. As it can be seen, the proposed framework works very well accross the different data sets and achieves performance of more than $99\%$, accross all the tested datasets. It is worth noting that these results are also general to complex spoofing cases such as the TEXBAT DS7 dataset, which most existing works have not successfully classified. Interestinly, in Fig.~\ref{confusion1} when the model is trained on simply spoofing scenarios, it still generalizes well to the complex spoofer in DS7. As such, our work provides superior performance compared to the state of the art\cite{Marata2025,10495074} (See Table \ref{TableComparison}) and it also enables the different spoofers to be identified by simply  using a single algorithm, thus, contributing to reduced computational complexity of receivers.
    
\section{Conclusions}
\label{conclusion}
In this paper, we proposed a novel meta-learning framework for RFF-based spoofer detection in GNSS receivers. We demonstrated that, by exploiting both the pre-correlation and post-correlation features, the performance of spoofing detection can be enhanced. This improvement is primarily due to the correlation present in the signal at different stages of the receiver. Furthermore, our proposed MAML framework enables receivers to detect spoofers with varying and unseen patterns, from simple to complex, thus providing a more general strategy for addressing spoofing threats. To advance this work, future research can focus on extending the framework to incorporate the ability to classify different forms of interference, allowing detection to be applied to other sources of disruption, such as jamming. In addition, future works could explore other advanced pretrained models.   
\section *{Acknowledgments}
The authors acknowledge the use of some AI tools to assist with minor language editing of this manuscript. 

\ifCLASSOPTIONcaptionsoff
  \newpage
\fi
\bibliographystyle{IEEEtran}
\footnotesize
\bibliography{IEEEabrv,ref.bib}
\end{document}